# Deep Learning Based Computer-Aided Systems for Breast Cancer Imaging : A Critical Review

Yuliana Jiménez-Gaona[a,b,c*], María José Rodríguez-Álvarez[b], Vasudevan Lakshminarayanan [c,d]

[a]Department of Chemical and Exact Sciences, Physicochemistry and Mathematics Section, Universidad Técnica Particular de Loja, Ecuador. [b] Instituto de Instrumentacion para la Imagen Molecular I3M, Universitat Politécnica de Valencia, Spain [c] Theoretical and Experimental Epistemology Lab, School of Optometry and Vision Science, University of Waterloo, Ontario, Canada [d] Department of Systems Design Engineering, Physics, and Electrical and Computer Engineering, University of Waterloo, Ontario, Canada
ydjimenez@utpl.edu.ec ,mjrodri@i3m.upv.es , vengulak@uwaterloo.ca

*Corresponding author at: Department of Chemical and Exact Sciences. Universidad Técnica Particular de Loja, Ap. 11-01-608, Loja- Ecuador. *E-mail address:* ydjimenez@utpl.edu.ec

**Abstract**

This paper provides a critical review of the literature on deep learning applications in breast tumor diagnosis using ultrasound and mammography images. It also summarizes recent advances in computer-aided diagnosis (CAD) systems, which make use of new deep learning methods to automatically recognize images and improve the accuracy of diagnosis made by radiologists. This review is based upon published literature in the past decade (January 2010 - January 2020). The main findings in the classification process reveal that new DL-CAD methods are useful and effective screening tools for breast cancer, thus reducing the need for manual feature extraction. The breast tumor research community can utilize this survey as a basis for their current and future studies.

## 1. Introduction

Due to the anatomy of the human body, women are more vulnerable to breast cancer than men. Breast cancer is one of the leading causes of death for women globally [1-4] and is a significant public health problem. It occurs due to the uncontrolled growth of breast cells. These cells usually form tumors that can be seen from the breast area via different imaging modalities.

To understand breast cancer, some basic knowledge about the normal structure of the breast is important. Women's breasts are constructed by lobules, ducts, nipples, and fatty tissues (**Fig 1**) [5]. Normally, epithelial tumors grow inside the lobes, as well as in the ducts, and later form a lump [6] generating breast cancer.

Breast abnormalities that can indicate breast cancer are masses and calcifications [7]. Masses are benign or malignant lumps and can be described by their shape (round, lobular, oval, and irregular) or their margin (obscured, indistinct, and spiculated) characteristics. The spiculated masses are the particular kind of masses that have high probability of malignancy. A spiculated mass is a lump of tissue with spikes or points on the surface. It is suggestive but not diagnostic of malignancy. It's a common mammography finding in the carcinoma breast [8].

On the other hand, microcalcifications are small granular deposits of calcium and may show up in clusters or patterns (like circles or lines) and appear as bright spots in a mammogram. Benign calcifications are usually larger and coarser with round and smooth contours. Malignant calcifications tend to be numerous, clustered, small, varying in size and shape, angular, irregularly shaped, and branching in orientation [7, 9].

Breast cancer screening aims to detect benign or malignant tumors before the symptoms appear, and hence reduce mortality through early intervention [2]. Currently, there are different screening methods such as mammography [10], magnetic resonance imaging (MRI) [11], ultrasound (US) [12], and computed tomography (CT) [13]. That helps to visualize hidden diagnostic features. Out of these modalities, ultrasound and mammograms are the most common screening methods for detecting tumors before they become palpable and invasive [2,14,15,16]. Also, they may be utilized effectively to reduce unnecessary biopsies [17]. These two are the modalities that are reviewed in this article.

A drawback in mammography is that the results depend upon the lesion type, the age of the patient, and the breast density [18-24]. In particularm dense breasts that are “radiographically” hard to see exhibit low contrast among the cancerous lesions and the background [25, 26].

Due to limitations of this modality such as low sensitivity especially in dense breasts, other modalities are used also, e.g., US [12]. The US is a non-invasive, non-radioactive, real-time imaging, and provides high image resolution images [27]. However, all these techniques are operator-dependent, and the interpretation of the images requires expertise in radiology. Normally, the radiologists try to do a manual interpretation of the medical image by double mammogram reading to enhance the accuracy of the results [28]. But it is time-consuming and is highly prone to mistakes [3, 29].

Because of these limitations, different artificial intelligence algorithms are gaining attention, due to their excellent performance in image-recognition tasks. Different breast image classification methods have been used to assist doctors in reading and interpreting medical images, such as Machine learning (ML), Deep learning (DL), and Computer-aided diagnosis/detection (CAD) systems [8, 30-32]. The CAD goal is to increase the accuracy of breast cancer detection rates by increasing sensitivity, which will support radiologists in their diagnosis decisions [33-35]. Recently, Gao et al., [36] developed a CAD system for screening mammography readings demonstrating about 92% accuracy in the classification. Likewise, other studies [37, 38] used convolutional neural networks (CNN) for mass detection in mammography and ultrasounds [39-41].

In general, the DL-CAD system (DL-CAD) focuses on CNNs which is the most popular model used for intelligent image analysis and for detecting cancer provide good performance [42-44]. With CNNs, it is possible to automate the feature extraction process as an internal part of the network [40], thus minimizing human interference [36]. DL-CAD systems have added broader meaning with this approach, distinguishing it from traditional CAD methods [36].
The next-generation technologies based on the DL-CAD system solve problems that are hard to solve with traditional CAD [45]. These problems including learning from complex data [46], image recognition [47], medical diagnosis [48,49], image enhancement [50]. In using such techniques, the analysis images include preprocesing, segmentation (selection of Region of Interest -ROI), feature extraction, selection, and classification.

In this review, we summarize recent advancements and developments in new DL-CAD systems for breast cancer detection/ diagnosis using mammograms and ultrasound imaging and then describe the principal findings in the classification process. The following research questions were used as the guidelines for this article:

- ¿How the new DL-CAD systems provide breast imaging classification-related concerning the traditional CAD system?
- ¿Which artificial neural networks inside the DL-CAD systems give better performance in breast tumor classification?
- ¿What are the main DL-CAD architectures used for breast tumor diagnosis/detection?
- ¿What are the performance metrics used for evaluating DL-CAD systems?

# 2.Methodology

### *2.1 Flowchart of the review*

The systematic review process is following the flow diagram and protocol (Fig 2) given in [51].

We identified appropriate studies in PubMed, Medline, Google Scholar, and Web of Science databases; as well as conference proceedings for IEEE, MICCAI, and SPIE published between January 2010 and January 2020. The search was designed to identify all studies in which Digital Mammography (DM) and the US were evaluated as a primary detection modality for breast cancer, and were both used for screening and diagnosis. A comprehensive search strategy included free text and MeSH terms, such as: “breast cancer,” “breast tumor,” “breast ultrasound,” “breast diagnostic,” “diagnostic imaging,” “deep learning,” “CAD system,” “convolutional neural network,” “computer-aided detection,” “computer-aided diagnoses,” “digital databases,” “mammography,” “mammary ultrasound,” “radiology information” and “screening” was utilized.

#### *2.1.1 Inclusion criteria*

Articles were included if they assessed Computer-Aided Diagnosis (CADx) or Detection (CADe) for breast cancer, DL in breast imaging, Deep convolutional neural networks (CNNs), DL in mass segmentation and classification in both DM and US, Deep Neural Network architecture, transfer learning and feature-based method in the automated mammography breast density. From a review of the abstracts, we manually selected the relevant papers.

*2.1.2 Exclusion criteria*

Articles were excluded if the study population included other screening methods such as MRI, CT, PET (positron emission tomography), or if other machine learning techniques were used.

***2.2 Study Design***

There are four sections in the design diagram (**Fig 3**). Firstly, different mammography and ultrasound public digital databases were analyzed as input data of the DL-CAD system. The second section includes the preprocessing and postprocessing in the traditional and next-generation DL-CAD.

In the third part, full articles were analyzed to compile the successful CNNs used in DL architectures. Furthermore, the best evaluation metrics were analyzed to measure the accuracy of these algorithms. Finally, a discussion and conclusions about these classifiers are presented.

***2.2.1 Public Databases***

Normally, the DL models are tested using private clinical images or public available digital databases, used by researchers in the breast cancer area. Public medical images are increasing because most of the DL-CAD systems require a large amount of data. Thus, DL algorithms are applied on available digitized mammograms such us MIAS (Mammographic Image Analysis Society Digital Mammogram Database) [52], DDSM (Digital Database for Screening Mammography), IRMA(Image Retrieval In Medical Application) [53, 54] INbreast [55] and BCDR (Breast Cancer Digital Repository) [36, 56] as well as ultrasound (US) public databases BUSI (Breast Ultrasound Image dataset), DDBUI (Digital database for breast ultrasound image), OASBUD (Open Access Series of Breast Ultrasonic Data) from Oncology Institute in Warsaw-Poland and the private US collected datasets SNUH (Seoul National University Hospital, Korean) [40], Dataset A (collected in 2001 from a professional didactic media file for breast imaging specialists) [57] and Dataset B (collected from the UDIAT Diagnostic Centre of the Parc Taulí Corporation, Sabadell- Spain). These widely used datsets are listed in **Table 1**.

***2.2.2 DL-CAD focused on DM and US***

The CAD systems are divided into two categories. One is the traditional CAD system and the other is the DL-CAD system. In traditional CAD system, the radiologist or clinician defines features in the image, and there can be problems in recognizing the shape and density information of the cancerous area. DL-CAD systems on the other hand create such features by itself through the learning process [67].

Further, CAD systems can be broken down into two main groups: computer-aided detection (CADe) and computer-aided diagnosis (CADx). The main difference between CADe and CADx is that the first refers to a software tool that assists in ROI segmentation within an image [68], identifying possible abnormalities and leaving the interpretation to the radiologist [8]. On the other hand, CADx serves as a decision aid for radiologists to characterize findings from medical images identified by either a radiologist or a CADe system.

### 2.2.3 Preprocessing

It is known that the database characteristics can affect significantly the performance of a CAD scheme, or even of a particular processing technique. Also, it can develop a scheme yielding erroneous or confusing results [69] since radiological images contain noise, artifacts, and other factors that can affect medical and computer interpretations. Thus, the first step in preprocessing is to improve image quality, contrast, and removal noises.

#### *2.2.3.1 Image Enhancement*

The main purpose of image preprocessing is to enhance the image and suppress noise while preserving important diagnostic features [70, 71]. Preprocessing in breast cancer also consists of delineation of tumors from the background, breast border extraction, and pectoral muscle suppression. The pectoral muscles are a challenge in mammogram image analysis depending on the standard view used during mammography. Generally, mediolateral oblique (MLO) and craniocaudal (CC) views are used [72].

As noted, DM includes many sources of noise, which are classified as a high-intensity rectangular label, low-intensity label, and tape artifacts. The principal noise models observed in mammography are Salt and pepper, Gaussian, Speckle, and Poisson noise.

In the same way, US images suffer from noise such as intensity inhomogeneity, low signal-to-noise ratio, high speckle noise [78, 79], blurry boundaries, shadow, attenuation, speckle interference, and low contrast. Speckle noise reduction techniques are categorized in filtering, wavelet, and compound methods [12].

Thus, many traditional filters can be applied for removal noise, including Wavelet transform, Median filter, Mean filter, Adaptive median filter, Gaussian filter, and Adaptive Wiener filter [3,73-77]. Also, different traditional methos: Histogram equalization (HE) [151,152], Adaptive Histogram Equalization (AHE)[153] and Contrast limited adaptive Histogram Equalization (CLAHE) [154] can be used to enhancement the image.

Actually, Deep-CNNs [155] are gaining attention for improving the super resolution image (SRCNN): Multi-image super-resolution, Example-based super-resolution and (iii) Single-image super-resolution [161-162]. Among the most used algorithms for generating high-resolution (HR) imaging [156,157] are: nearest-neighbor interpolation [158], bilinear interpolation [159], bicubic interpolation [160].

#### *2.2.3.2 Image Augmentation*

Deep CNN depends on large datasets to avoid overfitting and is necessary for good DL model performance [80]. Thus, limited datasets are a major challenge in medical image analysis [81], and it is necessary to implement data augmentation techniques. There are two common techniques for increasing the data in DL, data augmentation and transfer learning/fine-tuning [82, 83]. As an example, a DL model that has been trained with data augmentation is Imagenet [67]. Another effective example of transfer learning can be found in Huynh et al. [38].

The image augmentation algorithms include basic image manipulations (flipping, rotations, geometric transformations, color space augmentations, kernel, mixing images, random erasing [84]) and DL (feature space augmentation, adversarial training, generative adversarial networks (GAN) [85], neural style transfer [86] and meta-learning [81]). These techniques increase the amount of data by pre-processing input image data by operations such as rotation, contrast enhancement, and noise addition and has been implemented by many studies [189-196].

*2.2.3.3 Image Segmentation*

It processing step plays an important role in image classification. Segmentation is the separation of ROI (pectoral muscle [87], lesions, masses, microcalcifications) from the background of the image. In the case of cancerous images, we need the lesion part and from it extracts its features.

In traditional CAD systems, the tasks of specifying ROI such as an initial boundary or lesions, are accomplished with the expertise of radiologists. The traditional segmentation task in DM can be divided into four main classes: (i) threshold-based segmentation; (ii) region-based segmentation; (iii) pixel-based segmentation; and (iv) model-based segmentation [3,72]. Otherwise, US image segmentation includes several techniques: threshold-based, region-based, edge-based, water-based, active contour and neural network learning- based techniques [89, 90].

The accuracy of segmentation affects the results of CAD systems because numerous features are used for distinguishing malignant and benign tumors (texture, contour, and shape of lesions). Thus, the features may only be effectively extracted if the segmentation of tumors is performed with great accuracy [70, 88]. That is why, researhers are using DL methods especially CNNs, because it methods have excellent results on segmentation task. Also,DL-CAD systems are independent of human involvement and are capable of modeling breast US and DM knowledge using constraints autonomously. Two strategies have been utilized to used full image size for training CNN ond DM and US instead of ROIs. 1) High resolution[182] images and 2) patch-level [183].As for example recent network architectures used as a segmentator region are YOLO [163], SegNet[164], UNet[165-166], GAN[180] and ERFNet[181] .

***2.2.4 Post -Processing***

*2.2.4.1 Image feature extraction and selection*

After segmentation, feature extraction and selection are the next steps to remove the irrelevant and redundant information of the data being processed. Features are characteristics of the ROI taken from the shape and margin of lesions, masses, and calcifications. These features can be categorized into texture and morphologic features [12, 91], descriptor, and model-based features [92], and help to discriminate benign and malignant lesions. Most of the texture features are calculated from the entire image or ROIs using the gray level value and the morphologic features focus on some local characteristics of the lesion.

The high numbers of features increase the computational cost and slow down the classification process. Feature selection techniques reduce the number of feature space for developing process accuracy and minimizing computation time by eliminating redundant, irrelevant, and noisy features [93].

There are some traditional techniques used for feature selection like searching algorithms, chi-square test, gain ratio, information gain, recursive feature elimination, and random forest [94]. In addition, other traditional techniques used for the feature extraction include principal component analysis (PCA), wavelet packet transform (WPT) [95, 96], grey level co-occurrence matrix (GLCM) [97], Fourier power spectrum (FPS) [98], Gaussian derivative kernels [99], and decision boundary features [100].

However, in some advanced classification methods, such as an artificial neural network (ANN) and support vector machine (SVM), the dimension of feature vectors not only highly affect the performance of the classification but also determines the training time of the algorithm [91]. Thus, DL models produce a set of transformation functions and image features directly from the data [92], whose main advantage is to carry the burden of designing the specific features and the classification. Extracting useful features and make a good selection of the features is a crucial task for DL-CAD systems. As for example some CNNs capable of to extract features have been presented by different authors [184-188].

### *2.2.5 Classification*

After the features have been extracted and selected, they are input into a classifier to categorize the ROI into malignant and benign classes. The commonly used classifiers include linear, ANN, Bayesian neural networks, Decision tree, SVM, Template matching [70] and CNNs.

Recently, the deep CNNs, which are hierarchical architectures trained on large-scale datasets, have shown stunning performance in object recognition and detection [101], which suggests that these could also improve breast lesion detection in both US and DM methods. Some researchers are interested in lesion [178,196-198], microcalcifications [199-200] and masess [201-202] classification in DM and US [64, 120-124, 170,173,175] based on CNN models.

#### *2.2.5.1 Deep Learning Models*

DL in medical imaging is mostly represented by a basic structure called CNNs [64, 102]. There are different DL techniques such as Generative Adversarial Models (GANs), Deep Autoencoders (DANs), Restricted Boltzmann Machine (RBM), Stacked Autoencoders (SAE), Convolutional Autoencoders (CAE), Recurrent Neural Networks (RNNs), Long Short-erm Memory (LTSM), Multi-scale Convolutional Neural Network (M-CNN), multi-instance learning convolutional neural network (MIL-CNN) [3]. DL techniques have been implemented to train neural networks in breast lesions detection, include an ensemble of CNN [68] and transfer learning [68, 83, 101, 103]. The ensemble method combines multiple models to get a better and more comprehensive generalized model [104], and transfer learning is an effective method to deal with relatively small datasets as in the case of medical images.

ANNs are composed of an input and output layer, plus one or more hidden layers as shown in **Fig 4.** In the field of breast cancer, three types of ANN are frequently used: Back-propagation neural network, Self-organizing map (SOM), and hierarchical ANNs.

To train an ANN with a back-propagation algorithm the flow in the forward direction is used. Then, the generated output is matched with the desired output and the error signal is generated in the case the outputs do not match. This error propagates in the backward direction, and weights are adjusted for error reduction. This processing is repeated until the error becomes zero or is a minimum [105].

#### *2.2.5.2 Convolutional Neural Networks*

Convolutional neural networks (CNN) are the most widely used when it comes to DL and medical image analysis. The CNN structure has three types of layers, convolution, pooling, and full connection layer, stacked in multiple layers [67]. Thus, CNNs structure is determine by differente parameters such us number of hidden layers, learning rate, activation function (RELU), pooling layer for feature map extraction, loss function (Softmax) and the fully connected layers for classification. As shown in **Fig 5.**

##### *2.2.5.2.1 CNN architectures*

The model's performance depends on the architecture and the size of the data. There are different CNN architectures that have been proposed: AlexNet [106], VGG-16 [107], ResNet [108], Inception (GoogleNet) [109] and DenseNet [110]. They are briefly described in **Table 2**. These networks have shown promising performance in recent work for image classification.

Also, there are some techniques for improving the CNNs performance such us Dropout, Batch normalization and Cross-validation. Dropout is a regularization method to prevent a CNN model from overfitting. Batch normalization layer speeds up training of CNNs and reduce the sensitivity to network initialization and Cross-validation is a statistical technique to evaluate predictive models by partitioning the original samples into a training, validation and testing sets. There are three types of validation: 1. Hold-out splits (training 80% and testing 20%) 2. Three-way data splits (training 60%, validation 20% and testing 20%) and 3. K-fold cross-validation (3-5 k-fold for large data set; 10 k-fold for small dataset), where data is split into k different subsets depending on their size [176].

### ***2.2.6 Evaluation Metrics***

Different quantitative metrics are used to evaluate the classifier performance of a DL-CAD system. These include accuracy (Acc), sensitivity (Sen), specificity (Spe), the area under the curve (AUC), F1 score and confusion matrix (**Table 3 and Table 4**) that are shown for different architectures used in breast cancer (**Table 5** for DM and **Table 6** for the US) [39, 125].

*The Receiver Operating Characteristic Curve (ROC)*: The ROC curve is a graph of operating points, which can be considered as a plotting of the true positive rate (TPR) versus a false-positive rate (FPR), derived from AUC. The TPR and the FPR are also called sensitivity (recall) and specificity, respectively. As define in **Fig 6.**

*AUC*: Provides the area under the ROC-curve and a perfect score has a range from 0.5 to 1. It gives the aggregate measure of all possible classification thresholds. A 100% correct classified version will have the AUC value 1 and it will be 0 if there is a 100% wrong classification [126].

*Accuracy*: It indicador determines how many *TP*, *TN*, *FP,* and *FN* were correctly classified. In detail, Acc is the proportion of cases correctly classified into the benign or malignant state. Thus, TP: when a positive sample is classified correctl; FP: when a positive sample is classified incorrectly; TN: when a negative sample is classified correctly; FN: when a negative sample is classified incorrectly [68]

*Sensitivity or TPR:* Is the correct identification of true positive benign cases.

*Specificity or TNR:* Is the correct identification of true negative cases.

Additionally, some other frequently used statistical performance evaluation measures.

*Precision*: how precise is the model based on the true positive predicted correctly from the predicted ones.

*F1 score:* is the harmonic mean of the precision and recall, and;

*Matthew's correlation coefficient (MCC):* are also calculated to provide an efficient assessment of a classifier. Statistical equations are shown in **table 4.**

# 3. Discussion and Conclusions

Considering that the breast tumor screening using DM has some consequences (higher number of unnecessary biopsies and ionizing radiation exposure endangers the patient's health [12]) and limitations (low specificity, high FP results, which imply higher recall rates and higher FN results [132]) the US is used as the second choice for DM. Thus, US imaging is one of the most effective tools in breast cancer detection, because it has been shown to achieve high accuracy in mass detection, classification [133], and diagnosis of abnormalities in dense breasts [12].

For the aboved mentioned reasons, we have considered addressed this review using the both kind of images (DM and US), focus on different DL architectures applied in breast tumor processing. Offering a general overview of CNNs, including their relation and efficacy in performing segmentation, feature extraction, selection and classification tasks [134].

Thus, in **Table 2** various DL architectures and their training strategies for detection and classification tasks have been discussed. Based on the most popular datasets, CNN seems to perform rather well as demonstrated by Samala et al. [113], Cao et al., Chiao et al., and Yap et al. [63, 120, 121]. Also, the studies [111-121] used several preprocessing and processing techniques for high resolution [203], data augmentation, segmentation and classification. According to the most commonly CNNs used are AlexNet, VGG, ResNet, DenseNet, Inception (GoogleNet), LeNet and UNet, which employ rencently python libraries for implementing CNNs such us Tensorflow, Caffe and Keras with differente hyper-parameters to training the network [176].

Most of these network architectures use a large data set, thus it is required to apply an augmentation technique to avoid overfiting and to have better performance in classification. In this sense the researchers mentioned in Table 2 [167,170] used transfer learning and esemble methods as data augmentation to improving the performance of the CNN network, reaching 89.86% of accuracy and 0.9578% of AUC in DM, and AUC of 0.68% in US images. Furthermore, Singh, V. K. et al. [173] shows that the results obtained with GAN for breast tumor segmentation outperform the UNet model, and the SegNet and ERFNet models yield the worst segmentation results on BUS images.

In addition, according to Cheng J. et al. [175] the DL techniques could potentially change the design paradigm of the CADx systems for several profits over the traditional CAD. The profits are: First, DL can directly extract features from the training data. Second, the feature selection process will be significantly simplified. Third, the three steps of feature extraction, selection and classification can be realized within the optimization of the same deep architecture. Thus, SDAE architecture can potentially address the issues of high variation in either shape or appearance of lesions/tumors. Besides, the studies [176-179] show that CNN methods can compare images from CC and MLO views and can improve the accuracy of detection and reduce the FPR.

Besides, different evaluation metrics are described in Tables 3 and 4 as performance corroboration of these techniques. As a result, Tables 5 and 6 describe different research were their authors have used a variety of datasets (**Table 1**), approaches, and performance metrics to evaluate CNN techniques in DM and US imaging. For example, better results were achieved in DM analysis by Al-Masni [127] with YOLO5 using DDSM data augmentation, Chougrad et al. [129] used Deep CNN (Inception V3) with DDSM and MIAS datasets. On the other hand, Kyung et al [86] introduced a DenseNet model to analyze private (BUSI and SNUH) US datasets. Byra et al. [130] achieve high accuracy with the VGG19 Deep CNN model using the ImageNet database. Similarly, Cao et al [120] attained accuracy the 96.89% with SSD+ZFNet and Han et al. [53] with GoogleNet reached 91.23% using a private dataset.

Thus, according to the Springer (http://www.springer.com), Elsevier (https://www.elsevier.com), and IEEE (http://www.ieeexplore.ieee.org) web sites, researchers have mostly utilized the MIAS and DDSM databases for the breast image classification research. The number of conference papers published for the DDSM and MIAS databases is 110 and 168, respectively, with 82 journal papers published on DDSM databases and 136 journal papers published using the MIAS database [5]. Some details about the strengths and limitations of these databases are discussed in Abdelhafiz, D. [176].

Furthermore, **Table 7** gives a brief overview of the new DL-CAD system approach and traditional methods for CAD diagnosis. Even though Deheeba et al. [135] present a good traditional wavelet neural network CAD system with high accuracy (93.67%), Debelee et al. [136] exceed this percentage using CNN + SVM DL-CAD system in DDSM (99%) and MIAS (97.18%) datasets. These works demonstrate that in most of the cases DL architectures outperformed traditional methodologies.

To conclude, the use of DL could be a promising new technique to obtain the main features for automatic breast tumor classification especially in dense breasts. Also, in medical image analysis using DL has proven to be better for researchers compared to conventional ML approach [149, 150]. It appears as though DL provides a mechanism to extract features automatically through a self-learning network, thus boosting classification accuracy. However, there is a continuing need for better architectures, more extensive datasets that overcome class imbalance problems and better optimization methods.

**Acknwoledgments:** VL would like to acknowledge support by a Discovery grant from the Natural Sciences and Engineering Research Council of Canada.

**Figure Captions**

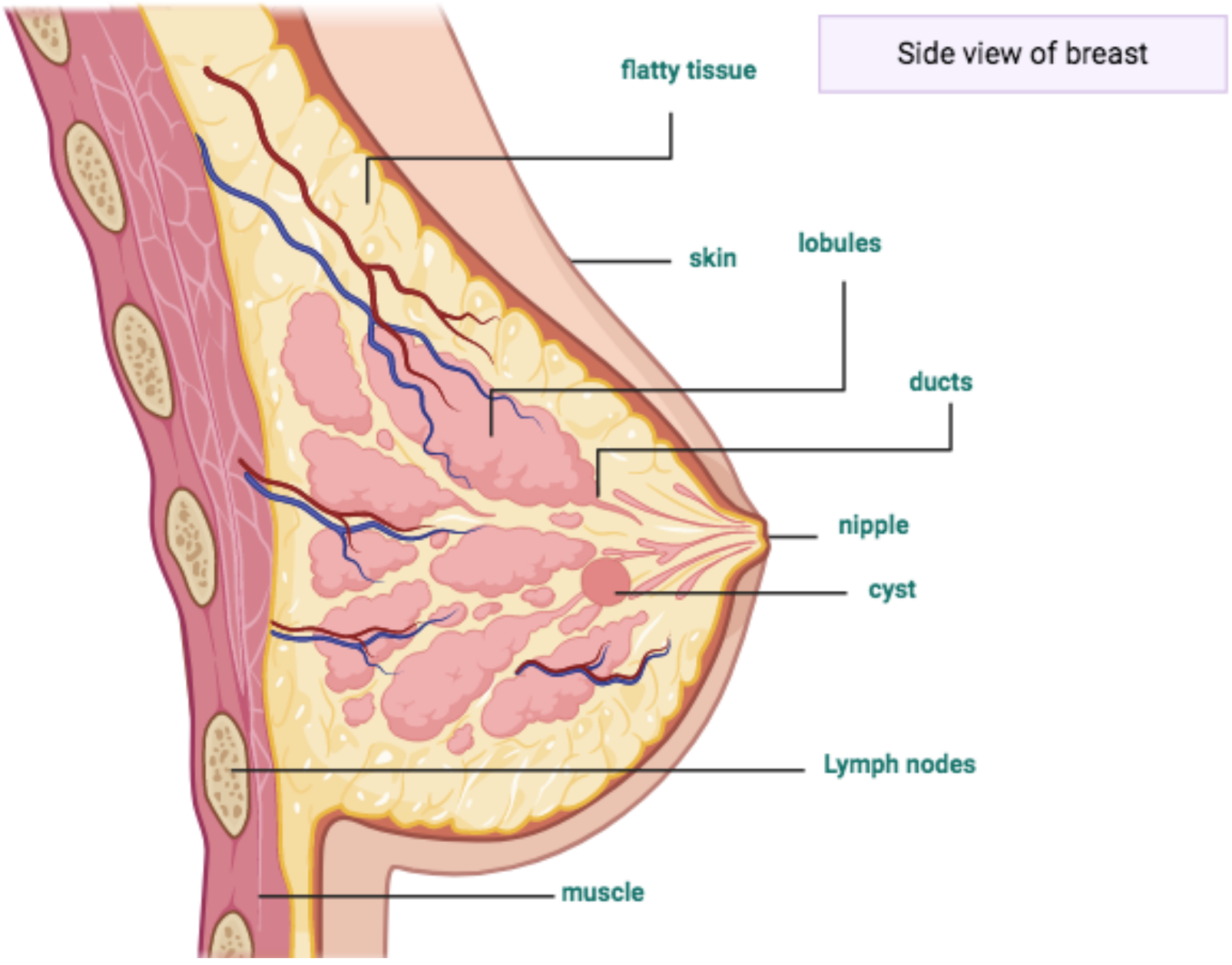

**Fig 1.** This scheme represents the anatomy of the woman breast. Inside the lobes are the zones where the epithelial tumors or cyst grow.  Designed by Biorender.

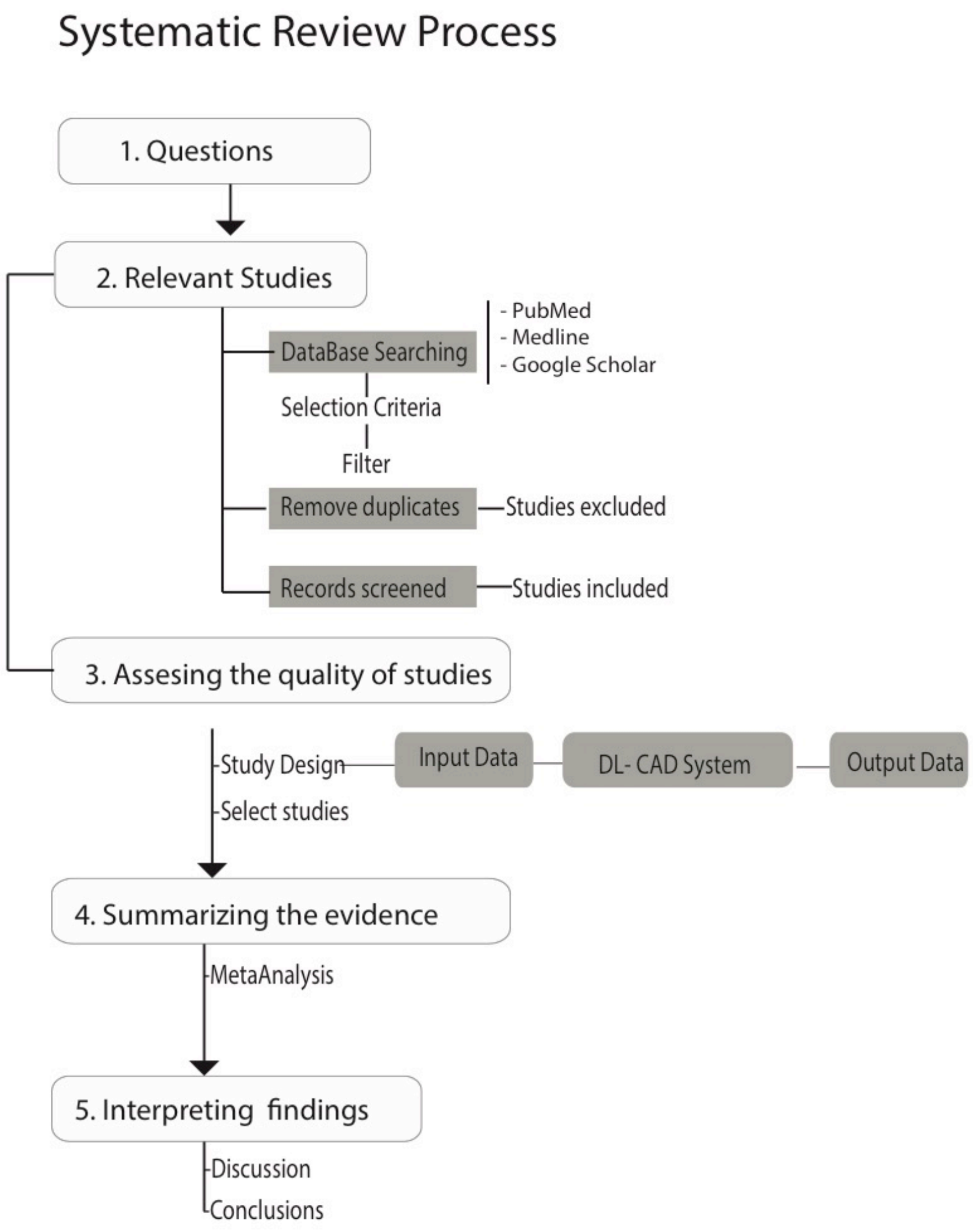

**Fig 2.** This Flowchart diagram represents the review process of articles in this paper

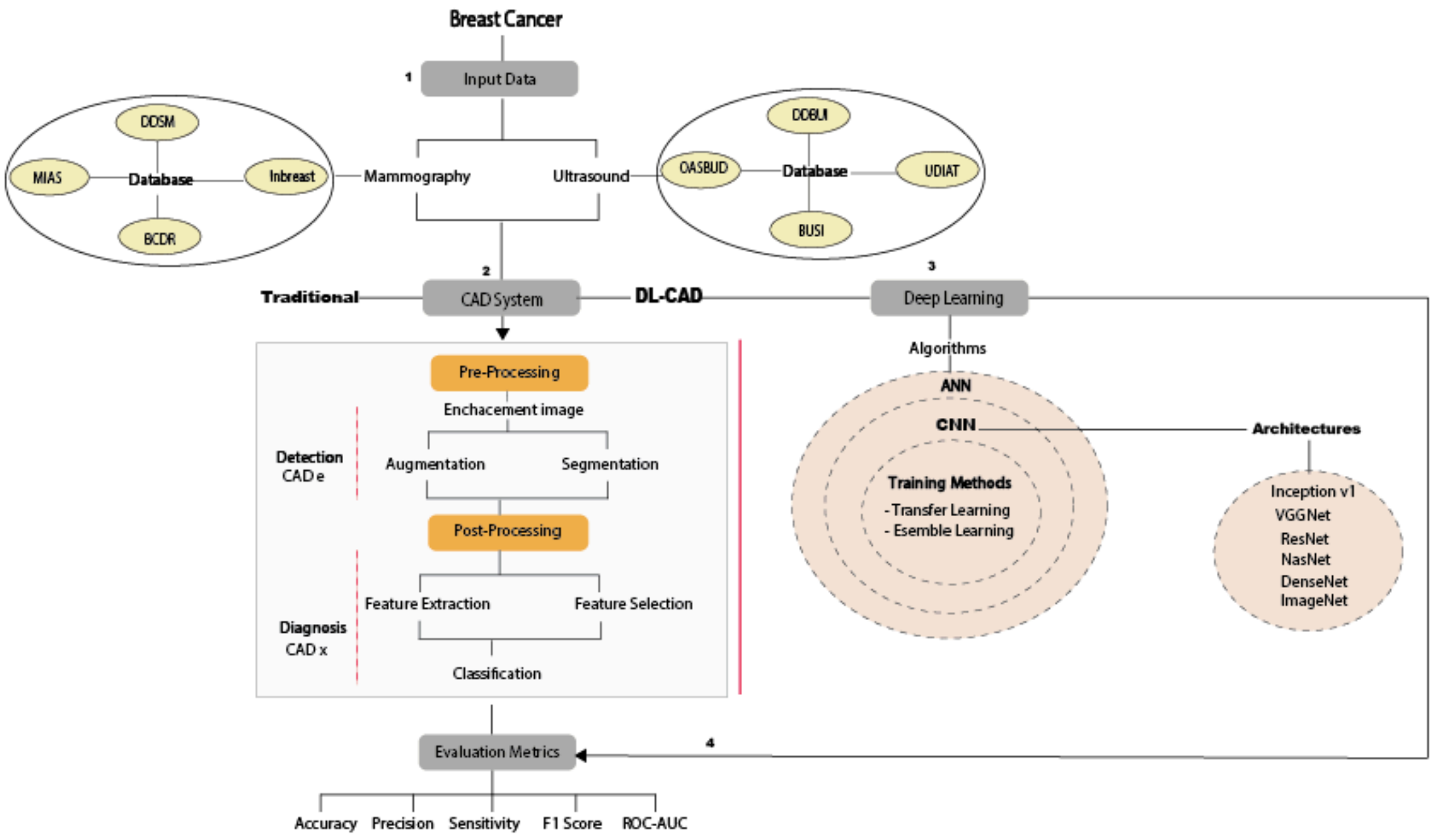

**Fig 3.** The general diagram is a flowchart which describes how an CAD system process can be used with DM and US images from public and private databases.. Normally, the CAD system consists of several stages such us segmentation, feature extraction-selection and classification. However, DL-CAD systems are based on CNN models and architectures to feature extraction-selection and classification with convolutional and fully connected layers automatically through a self-learning. Finally, CAD systems are validated by different metrics.

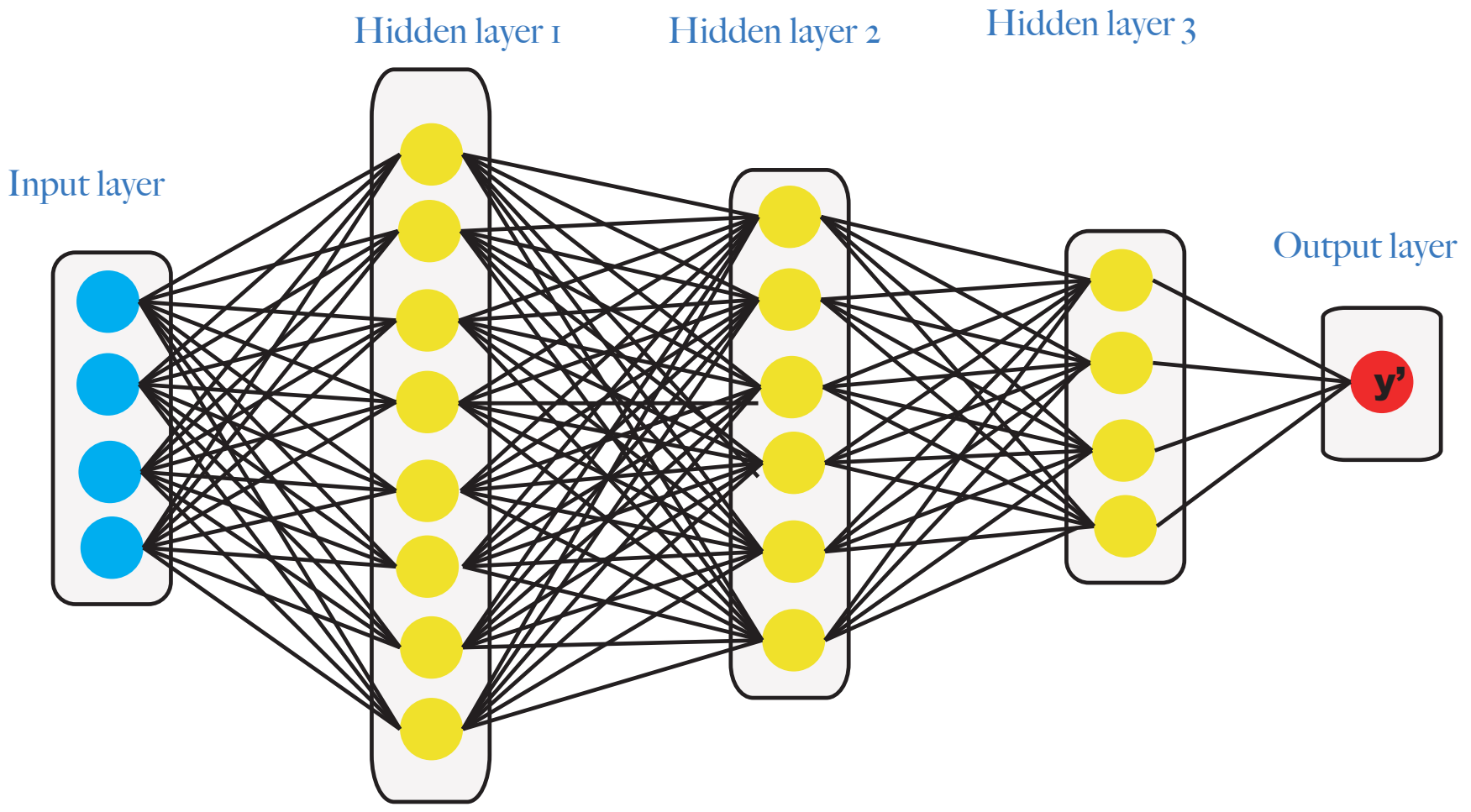

**Fig 4**. Artificial Neural Network (ANNs). It learns by processing images, each of which contains input, hidden and result layer.

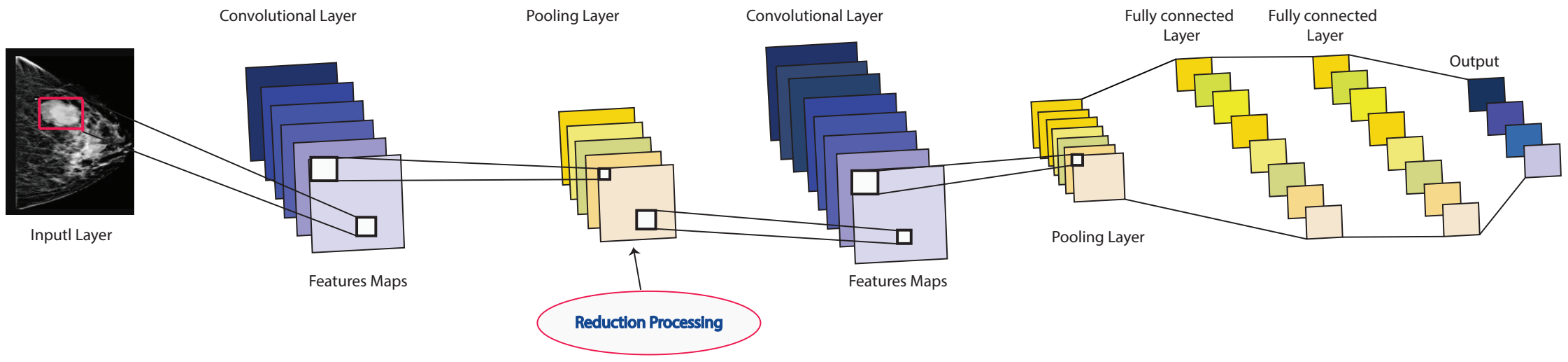

**Fig 5**. A feed-forward CNN network, where the convolutional layers are the main components, followed by a nonlinear layer (RELU), pooling layer for feature map extraction, loss function (Softmax) and the fully connected layers for classification. The output can be benign or malignant classes.

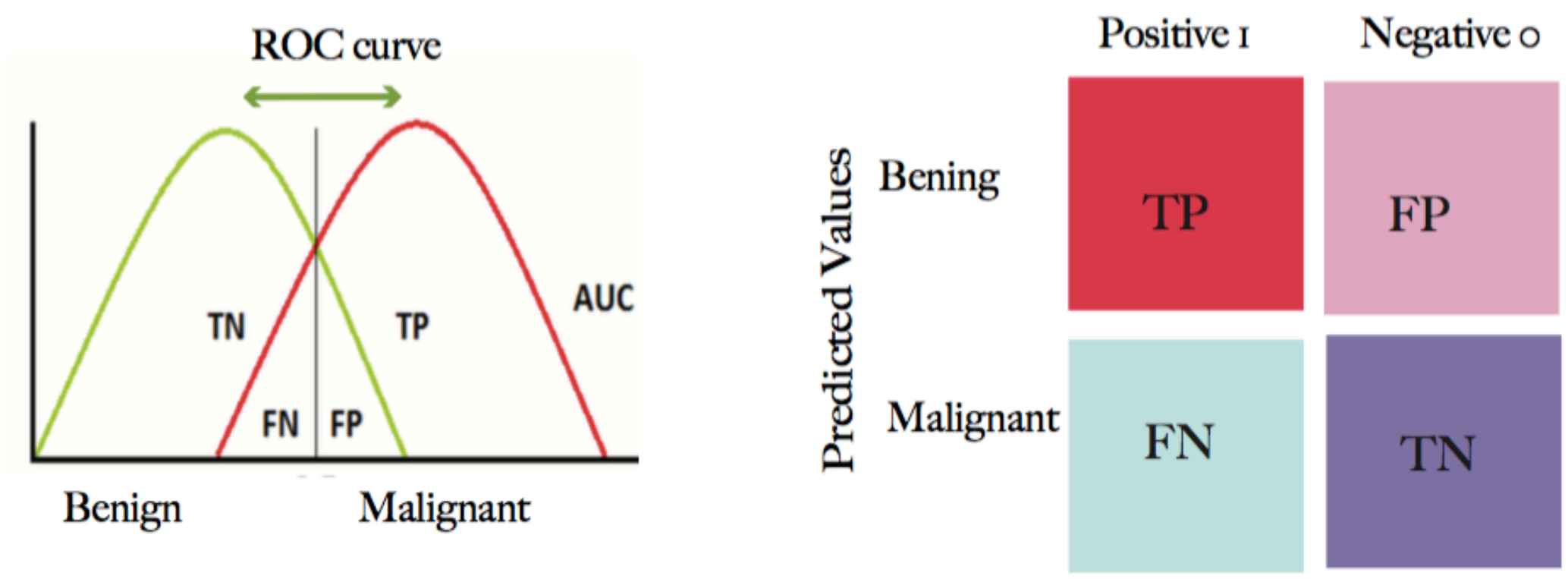

**Fig 6.** The confusion matrix regarding to the *Receiver Operating Characteristic Curve*. The number of images correctly predicted by the classifier is located on the diagonal. ROC curve utilizes TPR in the y-axis and the FPR fraction in the x-axis.

# TABLES

**Table 1**. Summary of public breast cancer databases, most commonly used in the literature.

| Type | Database | Annotations | Link | Author |
|---|---|---|---|---|
| Mammograms | DDSM | 2620 patients including MLO and CC | http://www.eng.usf.edu/cvprg/Mammography/Database.html | Jiao et al. [58] |
| | BCDR | 736 biopsy prove lesion of 344 patients including CC, MLO | https://bcdr.eu/ | Arevalo et al. [59] |
| | INbreast | 419 cases including CC, MLO 0f 115 patients | http://medicalresearch.inescporto.pt/breastresearch/index.php/Get_INbreast_Database | IMoreira et al. [60] |
| | Mini-MIAS | 322 digitized MLO images of 161 patients | http://peipa.essex.ac.uk/info/mias.html | Peng et al. [61] |
| Ultrasound | BUSI | The dataset consist in 600 female patients. The 780 images including 133 normal images without masses, 437 images with cancer masses, and 210 images with benign masses | https://scholar.cu.edu.eg/?q=afahmy/pages/dataset | Dhabyani et al. [62] |
| | DDBUI | 285 cases and 1132 images in total. | https://www.atlantis-press.com/proceedings/jcis2008/1735 | Tian et al. [63] |
| | Dataset A | Private dataset with 306 (60 malignant and 246 benign) images. | goo.gl/SJmoti | Yap et al. [64] |
| | Dataset B | Private dataset with 163 (53 malignant and 110 benign) images. | | Prapavesis et al. [57] |
| | SNUH | Private dataset with a total of 1225 patients, with 1687 tumors with biopsy-proven diagnosis were included in this study. | | Moon et al [40] |
| | OASBUD | 52 malignant and 48 benign masses | Zenodo repository (DOI: 10.5281/zenodo.545928)<br><br>website http://bluebox.ippt.gov.pl/~hpiotrzk | Piotrzkowska et al.[65] |
| | ImageNet | 882 US images, consisting of 678 benign and 204 malignant lesions. | http://www.image-net.org/ | Deng et al. [66] |

**Table 2.** Summary of CNN architecture information for breast imaging processing

| Reference | Model | Description | Way of training | Application |
|---|---|---|---|---|
| Krizhevsky et al. [111] | AlexNet | AlexNet is a deep CNN architecture and a winning model in the 2012 ImageNet Large Scale Visual Recognition Challenge [112]. Consist of 8 neural network layers, 5 convolutional. Some of which are followed by max-pooling layers, and 3 fully-connected layers with a final 1000-way softmax.<br><br>The Deep CNN was evaluated in Imagenet LSVRC-2010 dataset with a top-1 and top-5 error rates of 37.5% and 17.0%. And ILSVRC-2012 achieved a winning top-5 test error rate of 15.3%, compared to 26.2% | Trainig from dropout model to reduce overfitting in the fully-connected layers | ImageNet classification |
| Samala et al [113] | DL-CNN | (CAD) system for masses in digital breast tomosynthesis (DBT) volume using a Depp CNN with transfer learning from mammograms.<br>The performance of the trained DL-CNN and CNN for classification of the true microcalcifications and false positives in the test set is compared in terms of the area under the receiver operating characteristic (ROC) curve (AUC). The best AUC of 0.933 was obtained for the number of filters of (32, 32, 32, 32) with filter kernel sizes of 5, 5 and partial sum values of 18, 16 for the two convolution layers, respectively. The improvement was statistically significant ($p < 0.05$). | Training from CNN architecture with two hidden layers 12 and 8 node groups with filter kernel of 5 and 3 respectively. The classification performance of this CNN was compared with DL-CNN | Image detection in breast tomosynthesis from DM |
| Simoyan et al. [114] | VGG-VD | The very deep CNN models (VGG-VD16, VGG-VD19 [115]) were evaluated in ILSVRC-2014 | Training from Deep ConvNet architecture with 19 weight layers for largescale image classification | ImageNet classification |
| He et al. [116] | ResNet | These ResNet are easier to optimize, and can gain accuracy from considerably increased depth. An ensemble of these residual nets achieves 3.57% error on the ImageNet (ILSVRC 2015) test set. Evidence shows that the best ImageNet models using convolutional and fully-connected layers typically contain between 16 and 30 layers. | Training on the ImageNet dataset and evaluated by ResNet with a depth of up to 152 layers 8× deeper than VGG nets but still having lower complexity. | Imagenet recognition |
| Huang et al.[117] | DenseNet | Dense Convolutional Network (DenseNet), which connects layers in a feed-forward fashion way. Also, alleviate the vanishing-gradient problem, strengthen feature propagation, encourage feature reuse and substantially reduce the number of parameters. | Training on CIFAR-10, CIFAR-100, SVHN, and ImageNet with CNN with L layers and L connections. The network has L (L+1)/2 direct connections. For each layer, the feature-maps of all preceding layers are used as inputs, and its own feature-maps are used as inputs into all subsequent layers. | Object recognition |
| Szegedy et al. [27, 118] | Inception v5 | A deep CNN architecture for classification and detection in the ImageNet Large-Scale Visual Recognition Challenge 2014 (ILSVRC14) was proposed. | Training on Deep CNN, the main hallmark of this architecture is the improved utilization of the computing resources inside the network. | ImageNet clasification |
| Das et al [119] | VGGNet | BreakHist dataset with 58 malignat and 24 benign cases was evaluated with Deep CNN. The best accurate percentage was reached with 100x (89.06%). | Training from Multiple instances learning (MIL) architecture for CNN by designing a new pooling layer as multiple pooling layer (MPL) in torch 7. | Histopathology |
| Cao et al [120] | Deep CNN | Private dataset collected in Sichuan Provincial Hospital. Contains 577 benign and 464 malignant, | Training protocols for detection are Fast Region based CNN, Faster R-CNN, YOLO v3, SSD. YOLO and SSD perform significantly better than other methods.<br><br>Training protocols for classification are AlexNet, VGG, ResNet, GoogleNet, ZFNet and Densenet. Here DenseNet achieves best results than others methods. | US lesion dection and classification. |
| Chiao et al. [121] | Deep CNN | Private imaging dataset, collected from China medical University Hospital. Contains 307 images of ultrasound with 107 benign and 129 malignant. | Training from Mask R-CNN with ROI alignment to avoid the loss of spatial information.<br><br>Based on Faster R-CNN using región proposal network (RPN) to extract features. | Sonogram (ultrasound) lesion detection and classification. |
| Yap et al. [64] | LeNet UNet | CNNs have become an important technique in image analysis, particularly in image analysis. However, it has not been used in breast ultrasound lesion detection. For these reasons, this work studies the performance of CNNs in breast US lesion detection. | Deep learning approaches for breast US lesion detection in three different methods LeNet [122], U-Net [123] and transfer Learning [124], their performance is compared in two private datasets A and B | Breast lesion detection using US |
| Geras, K. et al. [167] | Multi-view DL-CNN | INBreast [58] and DDSM [60] databases were used in the research. As a result of this technique the size of features maps were greatly reduce, and the model achieved the AUC of 0.68%. | The network is trained jointly by stochastic gradient descent with back propagation [168] and data augmentation by random cropping [169]. Finally, a CNN aggressive with pooling layers was used for improving High-resolution [167]. | High Resolution, augmentation and DM classification. |
| Han, S. et al [170] | Google Net with Esemble learning | Dataset was built from 5151 patients' cases with a total of 7408 US breast images. Where 6579 were used as training set, 829 as test set (489 benign lesions and 340 malignant lesions). | CNN was training to differentiate malignant from benign tumors with optimal parameters as 10-fold cross validation. Data augmentation was necessary to make more robust the variability of breast tumor, employing the Caffe method. As results the mean accuracy reached 90.21% with 1 channel. | Data augmentation, detection and classification of breast lesions in US. |
| Dhungel, N. et al [172] | LeNet for CNN model s in cascad | INbreast dataset was used, with 115 cases and 410 images (116 cases contains benign or malign masses and the remaining ones do not contain any masses), from MLO and CC views. | For detection a cascade of DL methods (Fast R-CNN, Multi scale Deep Belief Network m-DBN, Random Forest) were used to select hypotheses based on Bayesian Optimization (Gaussian Mixture Model, GMM). | Detection, segmentation and classification of masses in DM. |
| | | | For segmentation a Deep structured output (Conditional Random Forest, CRF) learning was used.<br><br>For classification methodology two steps were proposed based on regression, 1. Pre training of CNN model and 2. Fine-tuning the pre-training data.<br><br>The final results showed that DL-CAD system is able to detect 90% of masses at one FPR per image, with segmentation accuracy of 85%, where the final classification (into benign or malignant) for the detected masses reached sensitivity of 0.98 and specificity of 0.7 | |
| Singh, V. K. Et al [173] | Generative Adversial Network (GAN) | The Mendeley database [174] was used, it contains 150 malignant and 100 benign tumors. Where the 70% was training set, the other 20% validation and the remaining 20%. | The experts first had manually segmented tumors. Then data augmentation was carried out by scale the images, gamma correlation, flip and rotate image. Finally, the segmentation was carried out with GAN learning. The metrics used to evaluate the performance were Dice and Intersection Over Union (IoU), it achieves scores of 93.76% and 88.82% respectively. | Breast tumor segmentation and classification in US images. |
| Cheng, J. Z. [175] | Stacked denoising auto-encoder (SDAE) based - CADx | The LIDC private database form Taipei Veterans General Hospital Taiwan was used, with 520 breast sonograms (275 benign and 245 malignant lesions) | A CNN model called OverFeat was used to classify nodules with the esemble method. The AUC performance reached around of 0.80. Furthermore, SDAE model was used to differentiate between distinctive types of lesions and nodules. | Breast lesion /nodules diagnosis in US images |

**Table 3**. Confusion matrix for a binary classifier to distinguish between two clases bening and malignant; where TP: is the number of True Positives; FN; False Negative; FP: False positive; TN: True Negative; TPR (true positive rate); FPR(false positive rate).

| Classes | Predicted classes | | Equation |
|---|---|---|---|
| | $C_1$ | $C_2$ | |
| $C_1$ (Bening) | TP | FN | $TPR=\left(\frac{TP}{TP+FN}\right)$ |
| $C_2$ (Malignant) | FP | TN | $FPR=\left(\frac{FP}{FP+TN}\right)$ |

**Table 4.** Validation assessment measures.

| Model | Equation |
|---|---|
| *Accuracy* | $Acc = \left(\frac{TP+TN}{TP+TN+FP+FN}\right)$ |
| *Sensitivity* | TPR **or** $Sen = \left(\frac{TP}{TN+FN}\right)$ |
| *Specificity* | TNR or $Spe = \left(\frac{TN}{TN+FN}\right)$ |
| *Precision* | $\mathrm{Pr}ecision = \left(\frac{TP}{TP+FP}\right)$ |
| *F1 Score* | $F1score = 2x\left(\frac{precision \times recall}{precision + recall}\right)$ |
| *MCC* | $MCC = \frac{TPxTN - FPxFN}{\sqrt{(TP+FP)(TP+FN)(TN+FP)(TN+FN)}}$ |

**Table 5.** Quantitative indicators used to evaluate the performance between different CNN architectures in DM datasets.

| Reference | Database | Deep CNN Model | ACC % | SEN % | SPEC % | Precision % | F1 Score % | AUC % |
|---|---|---|---|---|---|---|---|---|
| Al-Masni et al [ 127] | DDSM with 600 mamograms | CNN YOLO5- Fold cross validation in both datasets.<br>CAD mass detection<br>CAD mass classification | 96.73<br>85.52 | 93.20 | 78 | - | - | 87.74 |
| | DDSM augmentation with 2.400 mamograms | CAD mass detection<br>CAD mass classification | 97<br>99 | 100 | 94 | - | - | 96.45 |
| Ragab et al [126] | DDSM with 2620 cases | Depp CNN based Linear SVM<br>ROI Manually<br>ROI threshold | 79 | 76.3 | 82.2 | 85 | 80 | 88 |
| | CBIS- DDSM with 1644 cases | | 80.5 | 77.4 | 84.2 | 86 | 81.5 | 88 |
| | | SVM based médium Gaussian | 87.2 | 86.2 | 87.7 | 88 | 87.1 | 94 |
| Duggento et al. [128] | CBIS-DDSM | CNN | 71 | 84.4 | 62.4 | | | 77 |
| Chougrad et al. [129] | BCDR | Deep CNN Inceptionv3 | 96.67 | - | - | - | - | 96 |
| | DDSM | | 97.35 | - | - | - | - | 98 |
| | INbreast | | 95.50 | - | - | - | - | 97 |
| | MIAS | | 98.23 | - | - | -- | | 99 |

**Table 6.** Quantitative indicators used to evaluate different CNN architectures performance in US datasets.

| Reference | Database | Deep CNN Model | ACC % | SEN % | SPEC % | Precision % | F1 Score % | AUC % |
|---|---|---|---|---|---|---|---|---|
| Kyung et al [27] | BUSI SNUH | VGGNet Like | 84.57 | 73.65 | 93.12 | 89.34 | 80.74 | .91.98 |
| | | VGGNet 16 | 84.57 | 73.64 | 93.12 | 89.34 | 80.74 | 93.22 |
| | | ResNet 18 | 81.60 | 86.49 | 77.77 | 75.29 | 80.50 | 91.85 |
| | | ResNet 50 | 81.60 | 75.68 | 86.24 | 81.16 | 78.32 | 88.83 |
| | | ResNet 101 | 84.57 | 75,00 | 92.06 | 88.10 | 81.02 | 91.04 |
| | | DenseNet 40 | 85.46 | 79.05 | 90.48 | 86.67 | 82.69 | 93.52 |
| | | Dense1Net 12 | 86.35 | 77.70 | 93.12 | 89.84 | 83.33 | 92.48 |
| | | DenseNet 161 | 83.09 | 69.59 | 93.65 | 89.57 | 78.33 | 89.18 |
| Byra et al. [130] | ImageNet | VGG 19 combined with the Matching layer. | 88.7 | 0.848 | 0.897 | - | - | 93.6 |
| | UDIAT | | 84 | 0.851 | 0.834 | - | - | 89.3 |
| | OASBUD[115] | | 83 | 0.807 | 0.854 | - | - | 88.1 |
| Cao et al [120] | Private dataset consistin of 579 benign and 464 malignant cases. | Single Shot MultiBox Detector (SSD)300+ZFNet<br>YOLO<br>SSD300 +VGG16 | 96.89 | 67.23 | - | - | 79.38 | - |
| | | | 96.81 | 65.83 | - | - | 78.37 | - |
| | | | 96.42 | 66.70 | - | - | 78.85 | - |
| Han et al [53] | Private database with 5151 patients, with a total of 7408 US mages. 4254 benign and 3154 malignant lesions. | CNN-based GoogleNet | 91.23 | 84.29 | 96.07 | | - | 91 |
| Shan et al [131] | Private database, contains 283 breast US images, collected by the Second Affiliated Hospital of Harbin Medical University (Harbin, China)<br>Among them, 133 cases are benign and 150 cases are malignant. | ANN | 78.1 | 78 | 78.2 | - | - | 82.3 |

**Table 7.** DL- CAD system vs traditional methods

| Reference | Application | Method | Dataset | ACC % | SEN % | SPEC % | AUC % | Error % |
|---|---|---|---|---|---|---|---|---|
| Dheeba [135] | DM classification | Wavelet Neural Network | Private database consisting of 216 images of 54 patients taken in two different views CC and MLO. | 93.671 | 94.167 | 92.105 | 96.853 | 0.96853 |
| Trivizakis et al. [137] | DM classification | ML with transfer learning and feature-based with ImageNet and CNN architecture of 15 layers. | Mini MIAS 16 patients with 322 mammograms<br>DDSM 2,500 patients with 10,239 multi-view images including benign, malignant and normal cases. | 79.3 | - | - | 84.2 | - |
| | | | | 74.8 | - | - | 78.00 | - |
| Samala et al [138] | DM classification | Multi-task transfer learning framework to transfer knowledge learned from non-medical images by a Deep CNN. | ImageNet | 90 | - | - | 82 | - |
| Jadoon et al. [139] | DM Extraction | CNN discrete wavelet | IRMA with 2796 mammogram images, including 2,576 images from DDSM, 150 images from MIAS. | 81.83 | - | - | 83.1 | 15.43 |
| | | CNN curvelet transform, with SVM for classification | | 83.74 | - | - | 83.9 | 17.46 |
| Debelee et al. [136] | DM Extraction | CNN based SVM | MIAS (61 DM) | 97.46 | 96.26 | 100 | - | |
| | | | DDSM( 320 DM, 112 abnormal and the rest are normal) | 99 | 99.48 | 98.16 | - | - |
| | | MLP | MIAS | 87.64 | 96.65 | 75.73 | - | - |
| | | | DDSM | 97 | 97.40 | 96.26 | - | - |
| | | KNN-SVM | MIAS | 91.11 | 86.66 | 100 | - | - |
| | | | DDSM | 97.18 | 100 | 95.65 | - | - |
| Ahmed et al. [140] | DM detection | Deep CNN with 5 – fold cross validation | INbreast | 80.10 | 80 | - | 78 | - |
| Yuan Xu et al [141] | US image segmentation | CNNs 8-layer with convolutional and pooling layer 1-3, fully connected layer and softmax layer. Inputs of 128 x 128 | Private 3D breast US | 90.13 | 88.98 | - | - | - |
| Shan et al [142] | US image segmentation | ML methods: Decision tree | Private breast US consisting of 283 images with 133 cases are benign and 150 cases are malignant. | 77.7 | 74.0 | 82.0 | 80 | - |
| | | ANN | | 78.1 | 78.0 | 78.2 | 82 | - |
| | | Random forest | | 78.5 | 75.3 | 82.0 | 82 | - |
| | | SVM | | 77.7 | 77.3 | 78.2 | 84 | - |
| Gu et al [143] | 3D US image segmentation | Pre-procesing with morphological reconstruction minimizes the speckle noise and segmentation with region-based approach. | Private database with 21 cases, with masses prior of biopsy. Each case contains 250 slices. | 85.7 | - | - | - | - |
| Zhang et al | US image | Two-layer DL architecture for US (elastography) feature extraction and | The private dataset consisting of 121 female patients, with a total of 227 shear-wave elastography images, 135 of benign tumors | 93.4 | 88.6 | 97.1 | 94.7 | - |
| [144] | classification | classification, comprised of the point-wise gated Boltzmann machine (PGBM) and the restricted Boltzmann machine (RBM). Experimental evaluation was performed with five-fold cross validation on the dataset. | and 92 of malignant tumors. | | | | | |
| Almajalid et al. [145] | US image segmentation | DL architecture u-net, for breast ultrasound imaging. U-net is a convolutional neural network with two-fold cross validation designed for biology image segmentation with limited training data. | The private dataset contains 221 BUS images, collected from the Second Affiliated Hospital of Harbin Medical University in China. | 82.52 | 78.66 | 18.59 | - | - |
| Singh et al. [146] | US image classification | Classification technique by combining an unsupervised learning technique (i.e. fuzzy c-means clustering (FCM)) and supervised learning technique (i.e. back-propagation artificial neural network (BPANN)) | 178 B-mode breast US containing 88 benign and 90 malignant cases | 95.86 | 95.14 | 96.58 | 95.85 | - |
| Cheng et al [147] | US (sonograms) classification | DL architecture with Stacked denoising Autoencoder (SDAE) | 520 breast sonograms were scanned from 520 patients. The data involves 275 benign and 245 malignant lesions. | 82.4 | 78.7 | 85.7 | 89.6 | _ |
| Shi, et al [148] | US image classification | Deep polynomial network | A total of 200 pathology-proved breast US images (100 benign masses and 100 7 malignant tumors | 92.40 | 92.67 | 91.36 | - | - |